\documentstyle[multicol,epsf,aps]{revtex}
\begin{document}
\def\n{\nu}
\def\t{\tau} 
\def\D{\Delta}
\def\d{\delta}
\def\r{\rho}
\def\p{\pi}
\def\a{\alpha}
\def\g{\gamma}
\def\ra{\rightarrow}
\def\S{\sigma}
\def\b{\beta}
\def\e{\epsilon}

\title{On first-order phase transition in microcanonical and canonical
non-extensive systems}
\author{I.~Ispolatov\footnote[2] 
{corresponding author, Tel:(212)327-8860, Fax:(212)327-8507,
e-mail: slava@calif.rockefeller.edu} and E.~G.~D.~Cohen
}
\address{Center for Studies in Physics and Biology, Rockefeller University, 
1230 York Ave, New York,
NY 10021, USA.} 
\maketitle

\begin{abstract} 
\noindent  
Two examples of Microcanonical Potts models,
2-dimensional nearest neighbor and mean field, are considered
via exact enumeration of states and analytical asymptotic methods.
In the interval of energies corresponding to a first order phase transition,
both of these models exhibit a convex dip in the entropy vs energy plot  
and a region with negative specific heat within the dip. 
It is observed that in the nearest neighbor model the dip
flattens and disappears as the lattice size grows, while in the mean 
field model
the dip persists even in the limit of an infinite
system. If formal transitions 
from microcanonical to canonical ensembles and back are performed for an 
infinite but non-extensive  system ,
the convex dip in the microcanonical entropy plot 
disappears.

\medskip\noindent{PACS numbers: 05.70.Fh 64.10.+h 64.60.-i 82.60.Qr}

\end{abstract}

\section{Introduction}
Phase transitions in microcanonical ensembles (ME), particularly in 
finite systems or long-range interacting systems, which are both nonextensive,
have recently become a major subject of theoretical and computational study
\cite{hu1,hu2,wb,lw,bj,lb,gv,gr}. 
Although a lot has been done, in this work we bring up some points which 
may not 
be easy to deduce from the literature.

A distinct sign of a ME first-order phase transition  
is a convex ``dip'' \cite{hu2} or ``intruder'' \cite{gv} in an otherwise 
concave 
entropy plot (Fig.~1). 
\begin{figure}
\centerline{\epsfxsize=8cm \epsfbox{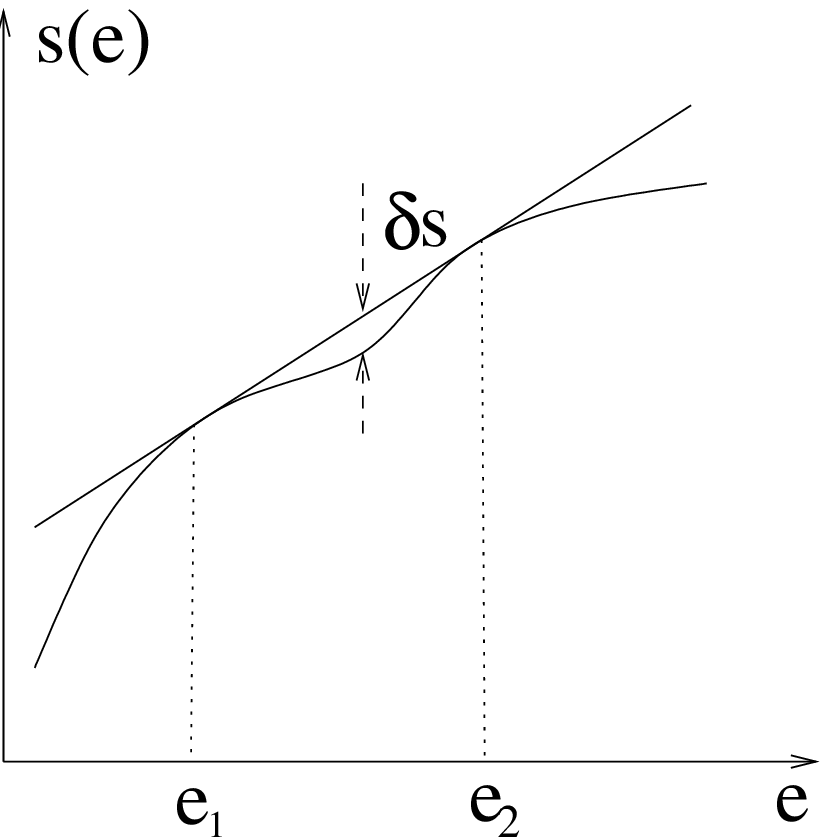}}
\noindent
{\small {\bf Fig.~1.}
Sketch of typical microcanonical specific entropy $s(e)$ as a function
of the energy per particle $e$
for a small system with a first-order phase transition  
A convex dip of the depth $\d s$ can be observed between $e_1$
and $e_2$.
}
\end{figure}
In the simple case of a system with a one-dimensional
thermodynamic space, which is considered in this paper,
a convex dip always includes
an interval between two points of inflection \cite{wb} 
where a second derivative of the specific entropy
is positive and the heat capacity is negative.
The convex dip in the entropy plot corresponds to the states of a two-phase 
coexistence; these states  are unstable in the
canonical ensemble (CE) and can not be accessed by equilibrium methods.
On the contrary, in the microcanonical ensemble, by tuning the energy, 
a system can be ``forced'' into these canonically unstable
two-phase coexistence states. Reasons for this are simple:
coexisting phases are separated by a boundary which carries a
certain entropic cost which is determined by the depth of the
dip $\d s$.  
If a system is small or the interaction is long-range so that 
the interfacial region occupies a significant part of the total volume, this
cost becomes important. Therefore if we put two weakly
interacting small clusters that are allowed 
to exchange energy with each other into a state near
the ``bottom'' of the dip, one of them will gain some energy and move 
close to the single-phase state 2, and
and the other will lose
some energy and move towards another single-phase state 1. As a result, the 
total entropy of two clusters will increase by $\approx 2\d s$.
If a system consisting of many such energy-exchanging clusters is put into a 
state where each cluster initially has energy between $e_1$ and $e_2$,
most probably it will arrange itself so that some cluster will be in state
$e_1$ and the remaining ones in $e_2$.
This scenario was frequently observed in  Monte-Carlo 
simulations of small Lennard-Jones clusters \cite{wd,lb1}, 
where the probability to find a cluster with an interface was very low.
On the contrary, in the ME one can tune the energy of a single cluster to
any state including those in the phase coexistence interval 
and therefore put a system into
a state with an interface even if it
is  entropically unfavorable.

Usually, as the system size increases and the contribution of the interface
to the total entropy becomes negligible, the convex dip gets 
shallower and in the limit of infinite system size becomes a straight line.
At this point phase coexistence states can be accessed by both ME and CE 
methods and two ensembles become equivalent.
However, if the interaction potential of the system is sufficiently long-range
(non-integrable at the large-$r$ limit), the relative interface contribution
does not disappear in the bulk limit. 
In this case a system with a first-order phase transition
in the ME can exhibit
an entirely different phenomenology than in the CE. 
For example, 
microcanonical gravitating systems may have regions with negative specific 
heat even in the infinite system size limit \cite{pr,lb}. 
Similar to the small cluster case, these states with negative 
specific heat can not be observed in the equilibrium 
CE since they are unstable.  

To illustrate both the short- and the long-range cases, we study 
the ferromagnetic 
$q$-state Potts model,
defined by Hamiltonian (\ref{H}).
\begin{equation}
H=-\sum_{i<j}\d_{s_i,s_j}, \;\; s_k=1,\ldots, q
\label{H}
\end{equation}
Two opposite limits are considered: Nearest Neighbor (NN)
interaction as an example of a system with short-range interactions 
and the Mean Field
(MF) theory as an example of a system with infinitely long-range interactions.
In the first case, the sum in (\ref{H}) runs only over nearest neighbor 
spins, in the second case, it includes all spins present in the system. In 
the MF model it is 
customary to introduce a $1/N$ prefactor to make the total energy scale 
as $\sim N$ in the $N\rightarrow\infty$ limit. 
With $q>4$ for NN in 2 dimensions  and $q>2$ for MF, both of these 
models have first-order phase transitions \cite{wu}. 
The $q=5$ NN model is studied via an exact microcanonical transfer matrix 
enumeration
on a two-dimensional square lattice of up to $6\times6$
sites. 
The $q\ge3$ MF models are studied both by
exact enumeration and continuous integration 
for a finite number of spins; an
analytical asymptotics for $s(e)$ is found in the infinite size limit. 

Given all the benefits of ME approach to first-order phase transitions 
outlined above, a natural question, ``what can be learned about
the ME properties from known CE behavior'' arises. 
To address this question, we
consider a typical example of a system with a convex entropy dip
and perform a transition from ME to CE and back.
It has been already stated \cite{hu2} that usually the ME data allow a
rather precise determination of all the CE properties but not the inverse.
The reason for this is that the inverse Laplace transform, which is performed
to obtain the ME entropy from the canonical free energy, 
is much less numerically stable than the direct one.
Moreover, in the last section of this paper we show that it is impossible 
to recover any information about the ME entropy convex dip from the 
corresponding
CE free energy if it is known only in the $N\rightarrow \infty$ limit.

\section{Exact enumeration of states for 2D Potts model}
Let us consider a 2D Potts model
on
a square lattice of the size $L\times L$.
Since we are looking for signs of a first-order phase transition, we
need to include into consideration at least the smallest possible value of 
$q$ for which such 
a transition exists in the bulk limit: $q=5$ \cite{wu}. A
microcanonical transfer matrix method developed by Binder \cite{bi} is used.
Cylindrical boundary conditions with open bonds at the top and the bottom
are utilized, a choice that will be justified
below. We first enumerate all 
possible states $|i\rangle,\: i=1,\ldots, q^L$ of a single 1-dimensional 
ring of spins, 
calculate their energies $E_i$, and
define a transfer matrix $\hat T$ with the elements $T_{ij}$ equal to the
energy of all bonds between two adjacent 
1D rings of spins that are in $|i\rangle$ and  
$|j\rangle$ states. Then we define the degeneracy $D_i^{1}(E)$ of a state 
$|i\rangle$ with the energy $E_i$ of a 1D 
spin ring, $D_i^{1}(E)=\d_{E,E_i}$. The degeneracy of state with the energy 
$E$ of a lattice consisting of 
a stack of $n$ rings 
where the state of the top ring is $i$ and the states of all lower rings
are summed over, can be recursively expressed:
\begin{equation}
\label{D}
D_i^{n}(E)=\sum_{j=1}^{q^L}D_j^{n-1}(E')\d_{E,E'+E_j+T_{ji}}.
\end{equation} 
We apply formula (\ref{D}) $N-1$ times and finally obtain for the 
degeneracy of the whole lattice 
\begin{equation}
\label{om}
D(E)=\sum_{i=1}^{q^L}D_i^n(E).
\end{equation} 
This approach allows us to reduce the number of operations to
$\sim L^2q^{2L}$ compared to $\sim q^{L^2}$ as for a straightforward exact
enumeration. However it requires enough memory to store the transfer matrix
which has $q^{2L}$ elements.
Applying a similar method to a system on a torus
(with completely periodic boundary conditions) will require storing
a degeneracy matrix $D_{ij}^{n}(E)$ with $2L^2q^{2L}$ elements 
and $\sim L^2q^3L$ operations.    
Hence we chose to limit our considerations to 
systems with cylindrical boundary conditions,
for which our computer resources allow us to study
$q=5$ spin systems on lattices 
with the base of the cylinder (length of the ring) up to 5.
The main quantity we are looking at is the specific microcanonical entropy 
defined as 
the logarithm of the density of states $\Omega(E)$ (which coincides with the 
degeneracy 
$D(E)$ if the
distance between energy levels is 1) divided by the number of
spins $N=L\times L$: $s(E)\equiv \ln(\Omega(E))/N$, 
with $E$ being the total energy of the system
For convenience we also use the intensive energy variable,
energy per spin $e=E/N$. 
For the Hamiltonian (\ref{H}) on the infinite 2D square lattice
$e$ varies between -2 and 0.
Plots of specific entropy $s(e)$ as a function of $e$ for 
various lattice sizes for $q=5$ are presented in Figure 2;
to be able to distinguish curves for different
system sizes,
we shifted plots for larger N upward. 
\begin{figure}
\centerline{\epsfxsize=8cm \epsfbox{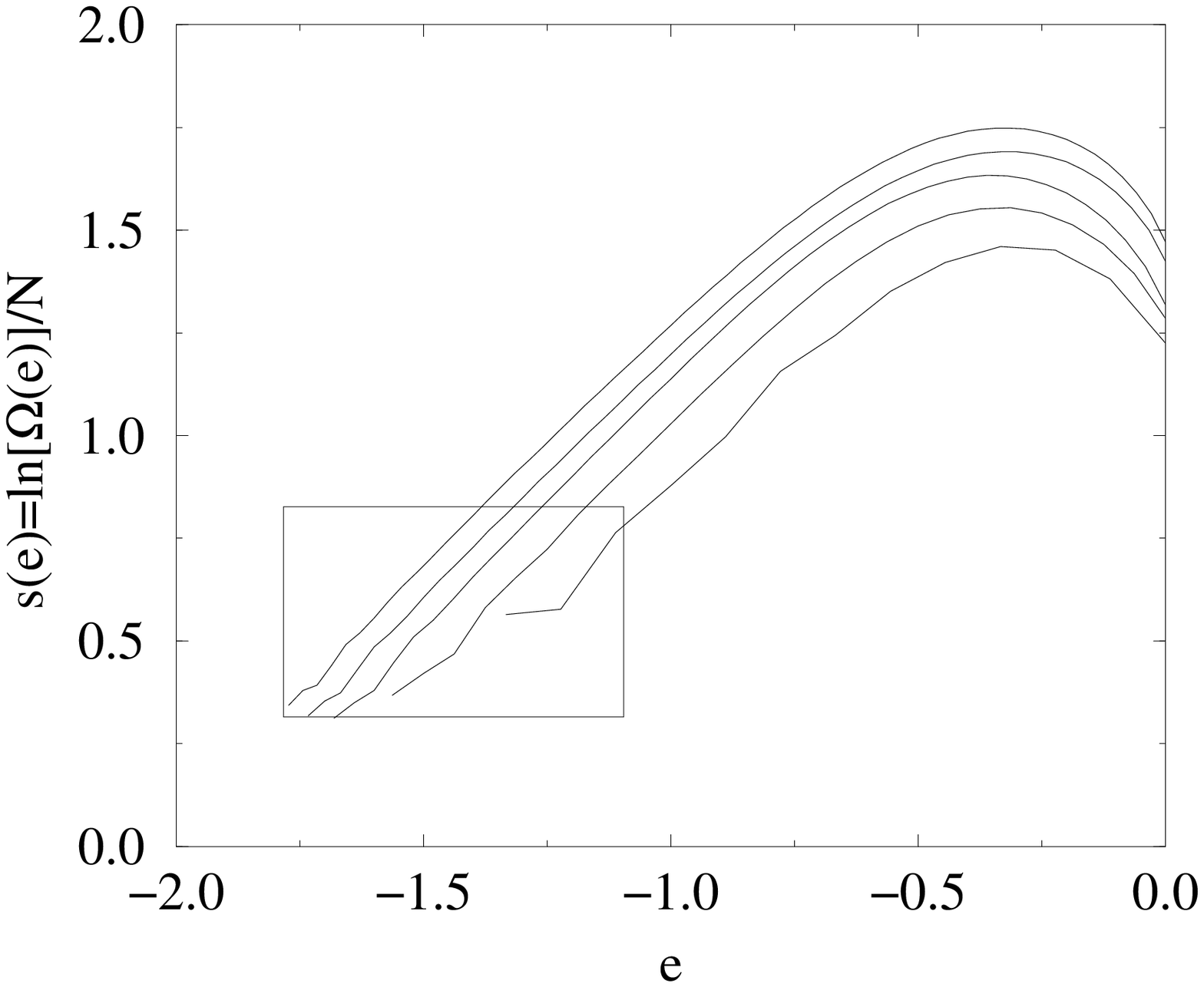}}
\noindent
{\small {\bf Fig.~2.}
Specific microcanonical entropy $s(e)\equiv \ln(\Omega(e))/N$ vs. 
energy per spin $e$ for $q=5$ for (top to the bottom)
7$\times$5, 6$\times$5, 5$\times$5, 4$\times$4,
and 3$\times$3 lattices. A box is drawn around convex dips in all 5 curves.
Ground states, which are separated from the first excited states
by $\Delta E=3$ energy gap, 
are not shown in all curves.
}
\end{figure}
Systems of all sizes (even with $N=9$ spins) exhibit definite 
signs of a phase transition: a dip between $e\approx-1.7$ and 
$e\approx -1.2$ 
can be clearly seen inside the box
in all 5 plots. The dip gets shallower as $N$ increases, which is consistent 
with reduction of the relative entropic cost of the interface for bigger 
systems.

\section{MF Potts models}
Let us consider next a Mean Field Potts (MFP) model defined by the Hamiltonian
\begin{equation}
H={-2\over N-1}\sum_{i=1}^q N_i(N_i-1).
\label{MF}
\end{equation}
Here  $N_i$ is a number of spins in the i{\it th} state, $0\leq N_i\leq N$,
$1\leq i \leq q$.
The coefficient $-2/(N-1)$, specific for a 2D square lattice, 
is introduced to make the MF Hamiltonian
consistent with the exact one (\ref{H}): with this coefficient 
the energy per spin $e$ in a perfectly ordered configuration is
also equal to -2.
One can view a MF model not only as an approximation
to a real system, 
but also as an independent model with non-decaying infinitely-long 
interactions. It is known \cite{wu} that in the canonical ensemble
the MF has a first-order phase transition for
$q\geq3$ and a second-order for $q=2$ (Mean Field Ising model).

The microcanonical partition function of the MF Potts model can be expressed
as
\begin{equation}
W({\tilde E})=\sum_{N_1=0}^N 
\ldots
\sum_{N_q=0}^N {N!\over \prod_{i=1}^q N_i!}\:
\d_{\sum_{i=1}^q N_i,N}\:
\d_{\sum_{i=1}^q N_i^2,{\tilde E}}.
\label{W}
\end{equation}
Here ${\tilde E}\equiv-E(N-1)/2+N$ (c.f. (\ref{MF})) is a rescaled energy of 
the system.
The two Kronecker $\d$ symbols are coming from the requirements that
the total number of particles and the total energy 
are equal to $N$ and ${\tilde E}$, respectively; 
the factorial ratio gives the  
statistical weight of the state.
To calculate the partition function,
for moderate $N$ ($N\leq 150$) we perform an exact enumeration
of states using integer arithmetics, while for larger $N$ ($150<N\leq 600$)
we use rational number calculations.
The results for the specific entropy $s(e)$ are presented in Figure 3.
\begin{figure}
\centerline{\epsfxsize=8cm \epsfbox{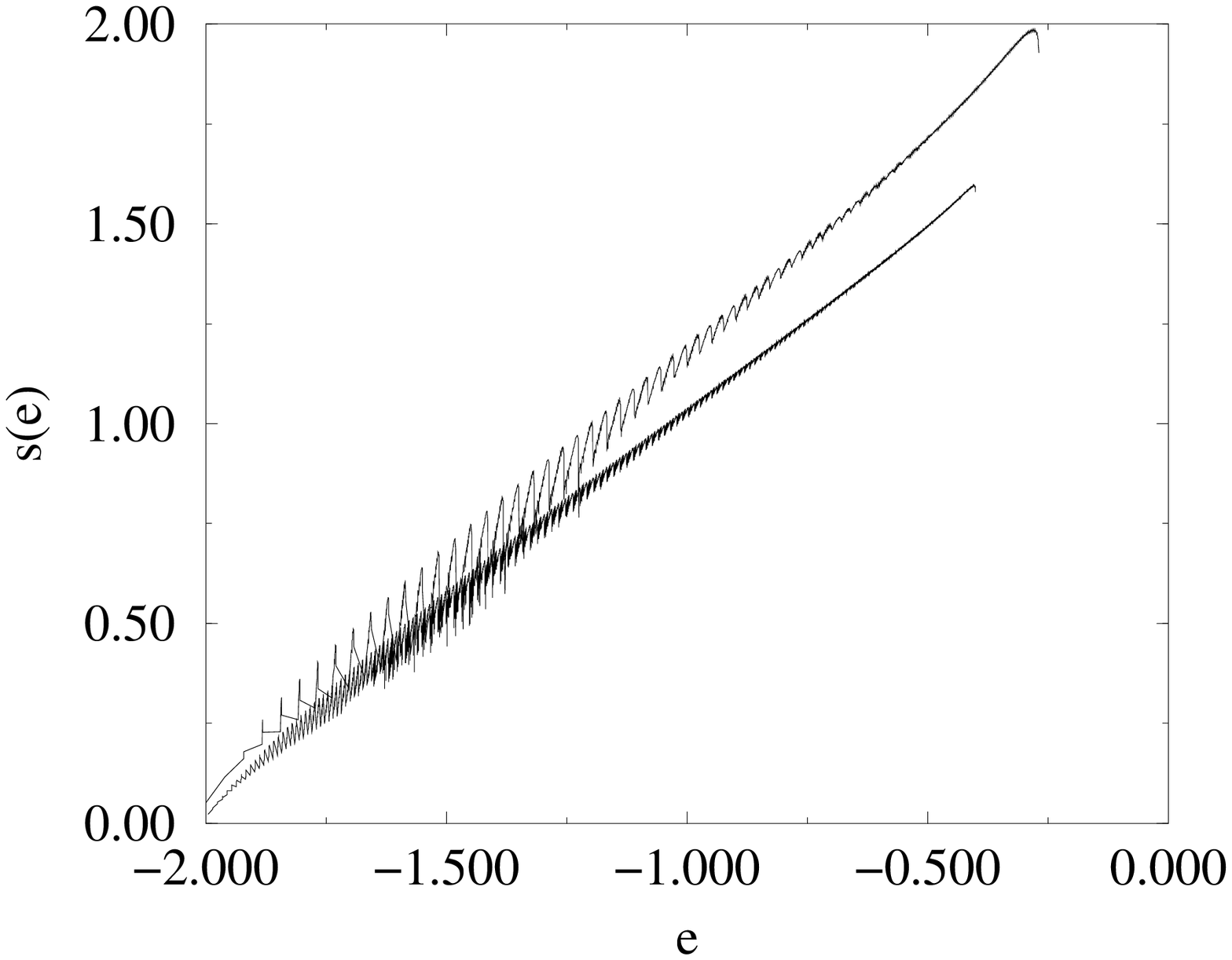}}
\noindent
{\small {\bf Fig.~3.}
Entropy $s(e)$ vs. 
energy per spin $e$ for $q=5$ and $N=500$
( line) and  $q=7$ and $N=100$ (lower line).
}
\end{figure}
Since energy states with high and low degeneracies are 
often observed to be  close to each other, these plots have a rather rugged 
structure.
It naturally leads to the assumption that only high-degeneracy states
contribute noticeably to a smooth coarsegrained entropy, which is defined
as the logarithm of the sum of the statistical weights of
a group of states with energies falling into a bin
$e-\D e/2, e+\D e/2$. The size of the bin $\D e$ must be large enough
to include several energy states.
Below we calculate the coarsegrained entropy and the leading term of 
its asymptotic behavior for $N\ra \infty$.

After replacing sums by integrals, factorials by continuous functions
via Stirling's formula, and
introducing the intensive variables $n_i\equiv N_i/N$, ${\tilde e}\equiv
{\tilde E}/N^2$, Eq.~(\ref{W}) becomes:
\begin{equation}
W({\tilde e})\approx N^{q-3}\int_0^1 \ldots \int_0^1  \d(\sum_{i=1}^q n_i-1)
\d({\sum_{i=1}^q n_i^2-{\tilde e}}) \exp(-N\sum_{i=1}^q n_i\ln n_i) dn_1
\ldots dn_q
\label{WC}
\end{equation}
Two integrations which eliminate the $\d$-functions are easily performed 
by introducing 
hyperspherical coordinates with the radius $r=\sqrt{\sum_{i=1}^{q}n_i^2}$ and
the polar axis $x_1=r\cos(\phi_1)$ 
along the $q$-dimensional
vector $(1/\sqrt{q},1/\sqrt{q},\ldots,1/\sqrt{q})$. The Jacobian
of this transformation is given by 
$r^{q-1}\prod_{i=1}^{q-1}[sin(\phi_i)]^{q-i-1}$ (see, for example, \cite{bp}).
\begin{equation}
W({\tilde e})\approx \theta(\tilde e-1/q)
{[N^2(\tilde e-1/q)]^{q-3\over2}\over 2 \sqrt{q}}
\int \exp(-N\sum_{i=1}^q n_i\ln n_i)\prod_{i=1}^{q}\theta(n_i) d^{q-2} \Omega,
\label{WCg}
\end{equation}

where $\theta(x)$ is the unit step function, and 
$d ^{q-2}\Omega$ is the remaining part of hyperspherical $q-1$ 
dimensional
angular differential
after the polar angle $\phi_1$ integration
has been carried out. The integration in (\ref{WCg})
runs over the remaining angular variables in the first quadrant
(only where all $n_i$ are non-negative) and is performed numerically.
For the simplest nontrivial case of $q=3$, the above expression transforms to
the following one-dimensional integral: (see Fig.~4)
\begin{figure}
\centerline{\epsfxsize=8cm \epsfbox{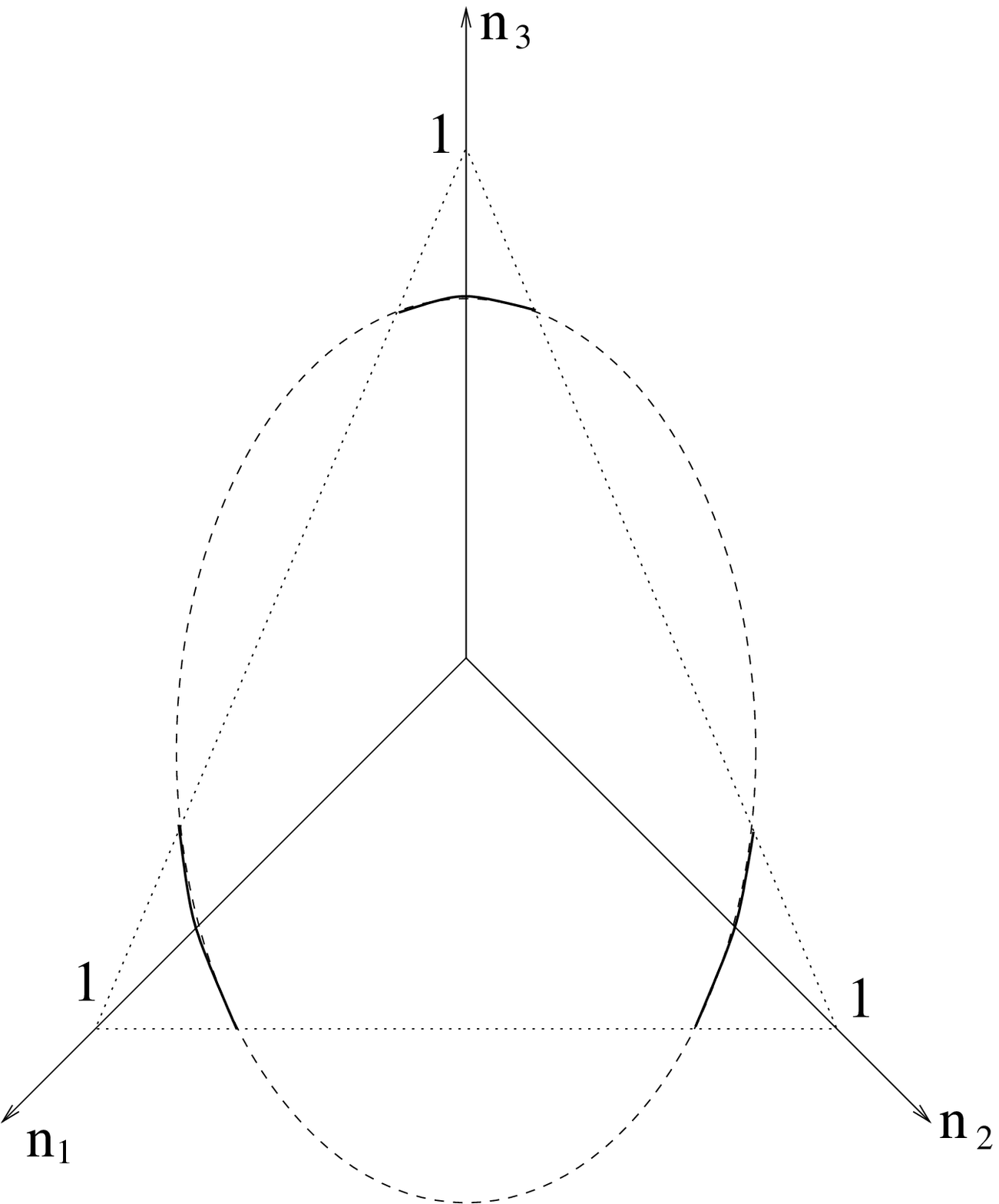}}
\noindent
{\small {\bf Fig.~4.}
Sketch of integration contour for (\ref{WC3})
for $q=3$ MF Potts model (3 solid lines segments). They lie on the
intersection of a circle with center 
$({1\over \sqrt{3}},{1\over \sqrt{3}},{1\over \sqrt{3}})$ and radius
$r=\sqrt{\tilde e -1/3}$ with the plane $n_1+n_2+n_3=1$. 
}
\end{figure}
\begin{equation}
W({\tilde e})\approx\theta(\tilde e-1/3) {1\over 2 \sqrt{3}}
\int_0^{2\pi} \exp(-N\sum_{i=1}^q n_i\ln n_i)\prod_{i=1}^{3}\theta(n_i) d\phi,
\label{WC3}
\end{equation}
where

\begin{eqnarray}
\label{ns}
\nonumber
&n_1={1\over 3}-\r({\cos\phi\over \sqrt{6}}+{\sin\phi\over \sqrt{2}})\\
\nonumber
&n_2={1\over 3}-\r({cos\phi\over \sqrt{6}}-{sin\phi\over \sqrt{2}})\\
&n_3={1\over 3}+2\r {\cos\phi\over \sqrt{6}}\\
\nonumber
&\r=\sqrt{\tilde e-1/3}.
\end{eqnarray}

In the limit of $N\rightarrow \infty$ the integral 
(\ref{WCg}) can be evaluated asymptotically for any $q$.
With the restrictions ${\sum_{i=1}^q n_i}=1$ and 
${\sum_{i=1}^q n_i^2}={\tilde e}$, the function $\sum_{i=1}^q n_i\ln n_i$
has $q$ identical 
maxima at which one $n_i$ (say, $n_1$) is large, and all 
the other $n_i$ are the same: 

\begin{equation}
n_1^*={1+\sqrt{(q{\tilde e}-1)(q-1)}\over q}, \;\;
n_2^*=\ldots = n_q^*={1-\sqrt{(q{\tilde e}-1)/ (q-1)}\over q}
\label{ma}
\end{equation}

To illustrate the dominance of the contributions from
the maxima (\ref{ma}), we present in Fig.~5 a
magnified segment of the plot of a non-coarsegrained (exact) 
$s({\tilde e})$ for $q=5$ and $N=500$ from Fig.~3, 
together with a sequence of points $(e_K,s_K)$ 
that are the
maxima defined by (\ref{ma})for integer $N_i$:

\begin{equation}
e_K={-2\over N(N-1)} [K^2+{(N-K)^2\over q-1}-N],\;\;
 s_K=-{K\over N}\ln{K \over N}-{N-K \over N}\ln{N-K\over N (q-1)},
\label{fw}
\end{equation}
where $K=N/q,\; N/q+1, \ldots, N$ labels the maxima.

\begin{figure}
\centerline{\epsfxsize=8cm \epsfbox{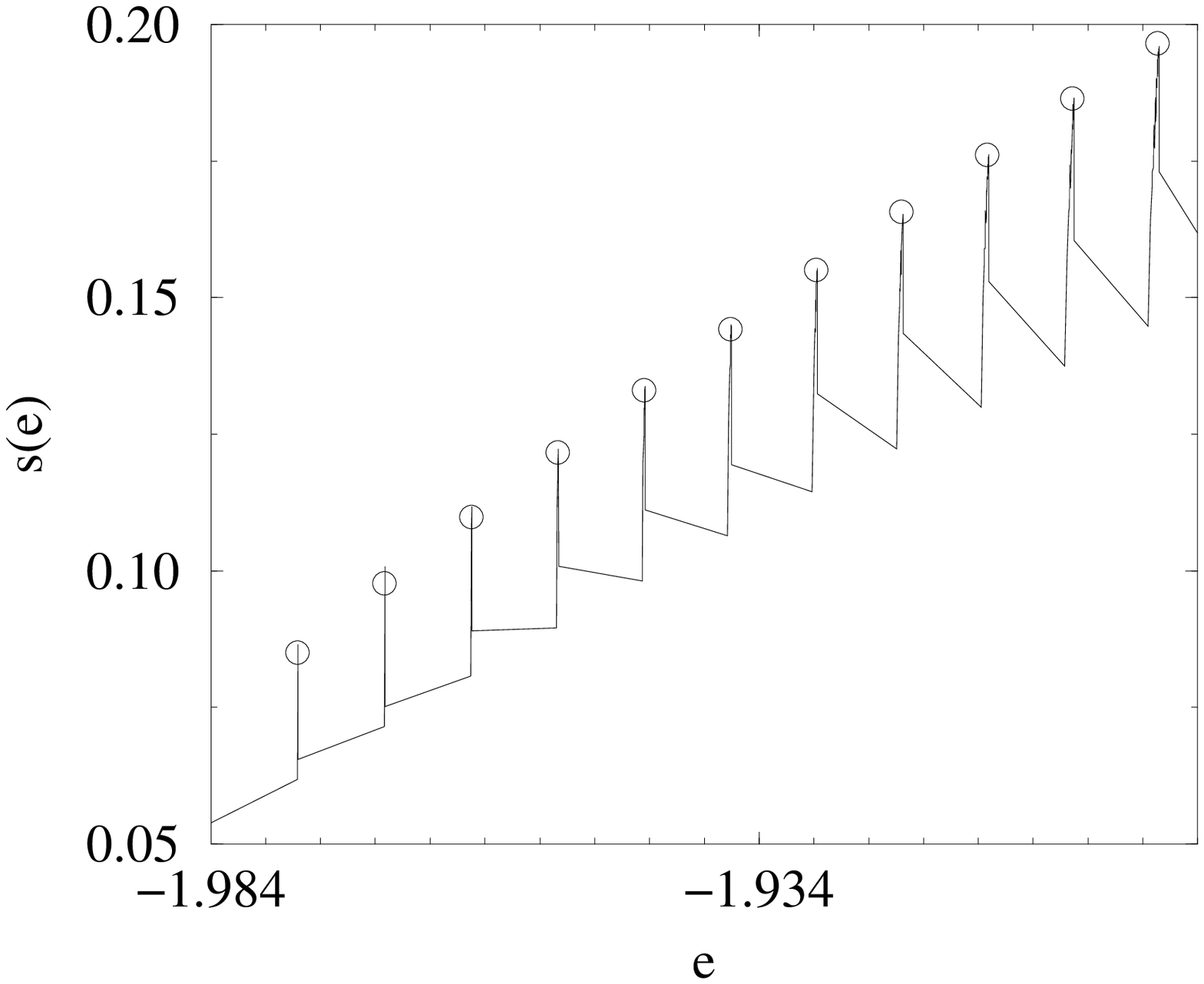}}
\noindent
{\small {\bf Fig.~5.}
Entropy $s(e)$ vs. 
energy per spin $e$ for $q=5$ and $N=500$
(solid line) and contribution from the
states defined by Eq.~(\ref{fw}) (circles).
}
\end{figure}

Evaluating the integrals in (\ref{WC}) by the steepest descent method 
around the maxima (\ref{ma}), we
obtain for the specific entropy:
\begin{equation}
s({\tilde e})= \ln(W({\tilde e}))/N=- \sum_{i=1}^q n_i^*\ln n_i^*+
{\cal O}({\ln(N)\over N}). 
\label{se}
\end{equation}

Now we proceed by looking at  signs of the convex dip in $s(e)$ and 
associated with 
it an interval with positive second derivative $s''(e)$, where the
microcanonical specific heat $C_v=-(s'(e))^2/s''(e)$ is negative. 
Generally, a deviation of the $s(e)$ plot from a straight line
is hardly visible, so we present the results through the
derivative $s'(e)$. This is done for a twofold purpose: first, 
a sign of the second derivative can be easily deduced from
the plots of $s'(e)$; second, recalling the definition of
the microcanonical
inverse temperature $\b(e)\equiv s'(e)$, the $s'(e)$ is of interest 
by itself. 
In Figure 6 we present the plot of $s'(e)$ for the $q=3$ MFP model 
for various
numbers of spins $N$ ranging from 50 to infinity. 
Curves for finite $N$ are 
produced by direct numerical evaluation
of the integral (\ref{WC}); knowledge of the $N\rightarrow \infty$
asymptotics (\ref{se}) helps to avoid real number overflow 
for large $N$.
The bottom line for
$N=\infty$ is the exact steepest descent result (\ref{se}).

\begin{figure}
\centerline{\epsfxsize=8cm \epsfbox{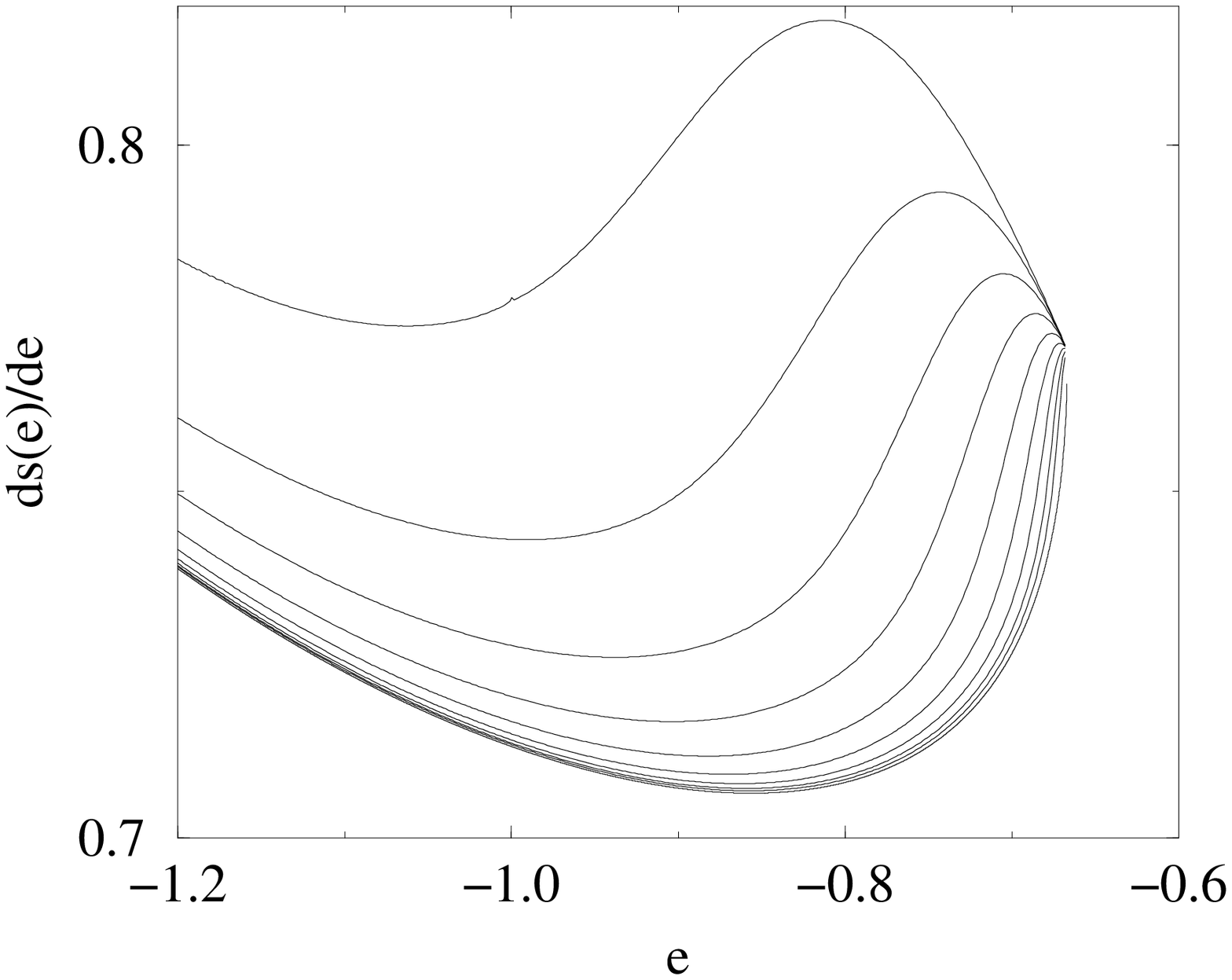}}
\noindent
{\small {\bf Fig.~6.}
Specific entropy derivative $s'(e)$ vs. 
energy per spin $e$ for $q=3$ and for 10 various $N$ 
(going from the up-most line down) $N=50, 100, 200, 400, 800,
1600, 3200, 6400, 12800, \infty$.
}
\end{figure}
For all system sizes $N$, $s'(e)$ approaches the same value in the 
high-energy limit,
$s'(e=-2/3)=3/4$ (c.f. (\ref{WC3}), (\ref{se}).
For any finite $N$, the second derivative of the  entropy
also approaches a finite negative value, $s''(e)\rightarrow -9/16$,
as $e\rightarrow-2/3$, but the interval near $e=-2/3$, where
$s''(e)$ is negative, shrinks as $1/\sqrt{N}$.

In Figure 7, $s'(e)$ is plotted 
for various $q$ ranging between 
2 and 100 for an infinite system.
\begin{figure}
\centerline{\epsfxsize=8cm \epsfbox{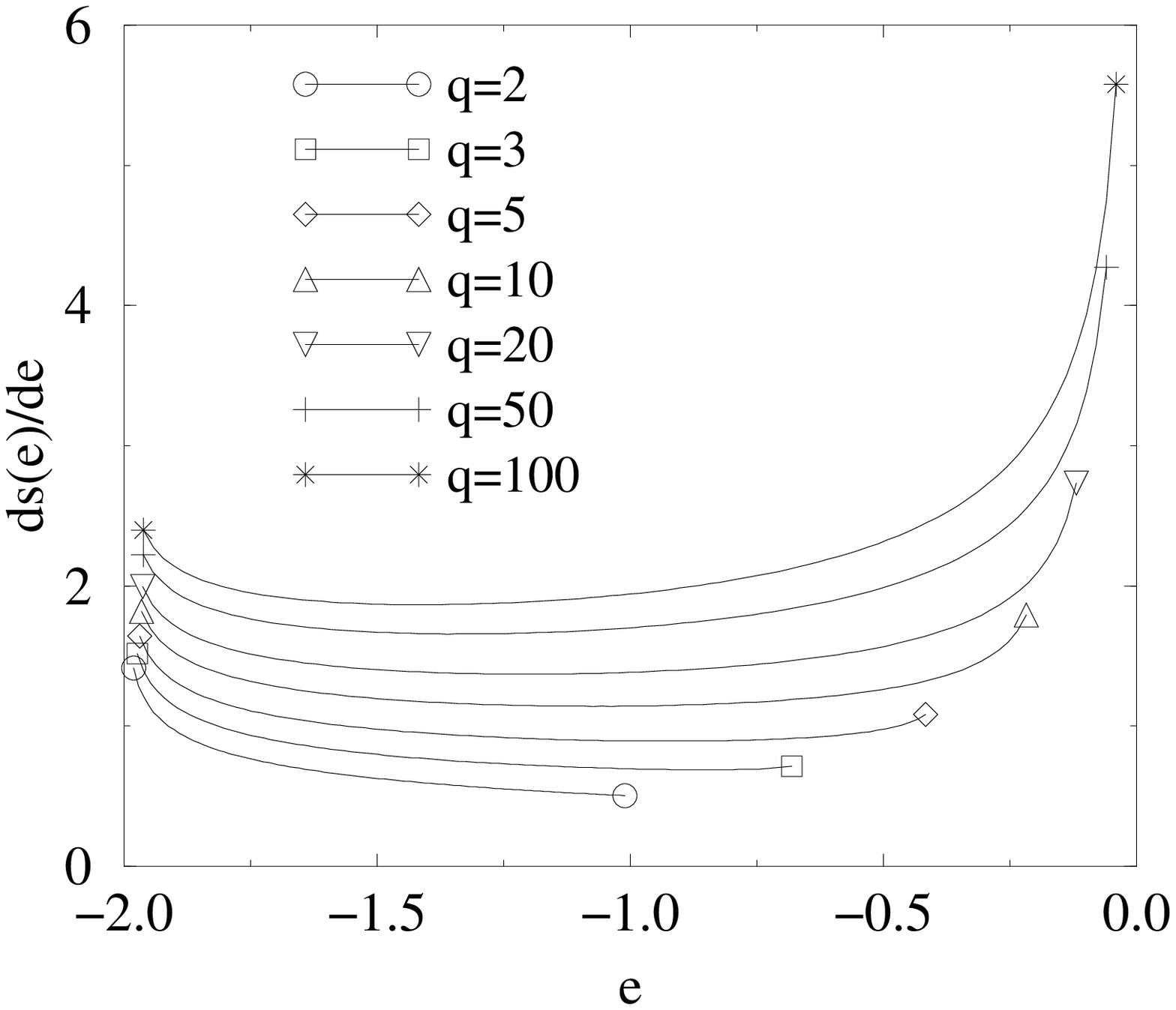}}
\noindent
{\small {\bf Fig.~7.}
Entropy derivative $s'(e)$ vs. 
energy per spin $e$ for $N=\infty$ and for 7 values of $q$ 
(going from the upper line down) $q=100, 50, 20, 10, 5, 3, 2$.
}
\end{figure}

For $q=2$, which corresponds to the Ising mean-field model,
$s''(e)$ is negative for all allowed energies.
This is consistent with the canonical ensemble results for the
MF Potts $q=2$ model, 
where the specific heat has finite limits on both sides of a 
second-order phase 
transition and experiences a discontinuous jump at the phase
transition point. 
The microcanonical inverse temperature $\beta=s'(e)$ at the point 
of phase 
transition $e=-1$, which is the high end point of the 
allowed energy interval $-2 \leq e \leq 2/q$,
is equal to its MF Potts canonical equivalent, 
$\beta_c=1/2$, given that the energy 
scale is
defined as in the Hamiltonian (\ref{H}).
For $q\geq 3$, a region of positive 
$s''(e)$ appears and gets wider with increasing $q$. 
At the upper boundary of the allowed 
energy interval, $e=-2/q$, the entropy derivative grows linearly with $q$,
$s'(e=-2/q)=q/4$.
 
Figures 6 and 7 indicate that MF Potts model does 
exhibit a convex dip in the entropy plot and an associated interval
with negative specific heat, even in the infinite system limit.
The reason for this is the distance-independent nature of interactions
in the MF Hamiltonian which makes phase segregation impossible. Hence
the entropic cost of phase coexistence scales linearly with the number of 
spins in the system and the depth of the convex dip in the entropy plot goes
to a finite limit as $N\rightarrow \infty$.

\section{connection to canonical ensemble}

In the previous sections we performed a microcanonical analysis 
of the  Potts model which is one of the most simple systems with a 
first-order phase transition. 
However, in general 
the canonical properties are much
better studied than the microcanonical ones. 
A question then arises: what can be learned about
the microcanonical behavior of a system which has a first-order
phase transition from its canonical solution?
A general answer is well-known: if a system is infinite and extensive, 
i.e. the interaction is not long-range, then microcanonical and 
canonical behavior are the same. However, the MF Potts model can be 
considered as having  infinitely long-range interactions and,
as we showed above, does have a negative specific heat even in the
infinite system size limit. Below we present a description of
what happens to the negative specific heat interval and the convex dip
in the entropy plot 
as a whole, if we formally attempt a transformation from the
microcanonical ensemble to the canonical ensemble (i.e. a Laplace 
transform of
microcanonical 
partition function) and then transform back from the canonical to the
 microcanonical ensemble 
(i.e. inverse Laplace transform of the canonical partition function). 
This procedure discussed below is of a rather general 
nature and can be applied to any system with a convex dip
in the microcanonical entropy plot; to the best of our knowledge, 
such a general 
description has yet not been presented in the literature.

To calculate the canonical partition function $Z(\b)$,
\begin{equation}
\label{z}
Z(\b)=N \int e^{N[s(e)-\b e]} de,
\end {equation}
let us consider an interval of energies where the specific entropy
$s(e)$ has a convex dip (Fig.~8):
\begin{figure}
\centerline{\epsfxsize=8cm \epsfbox{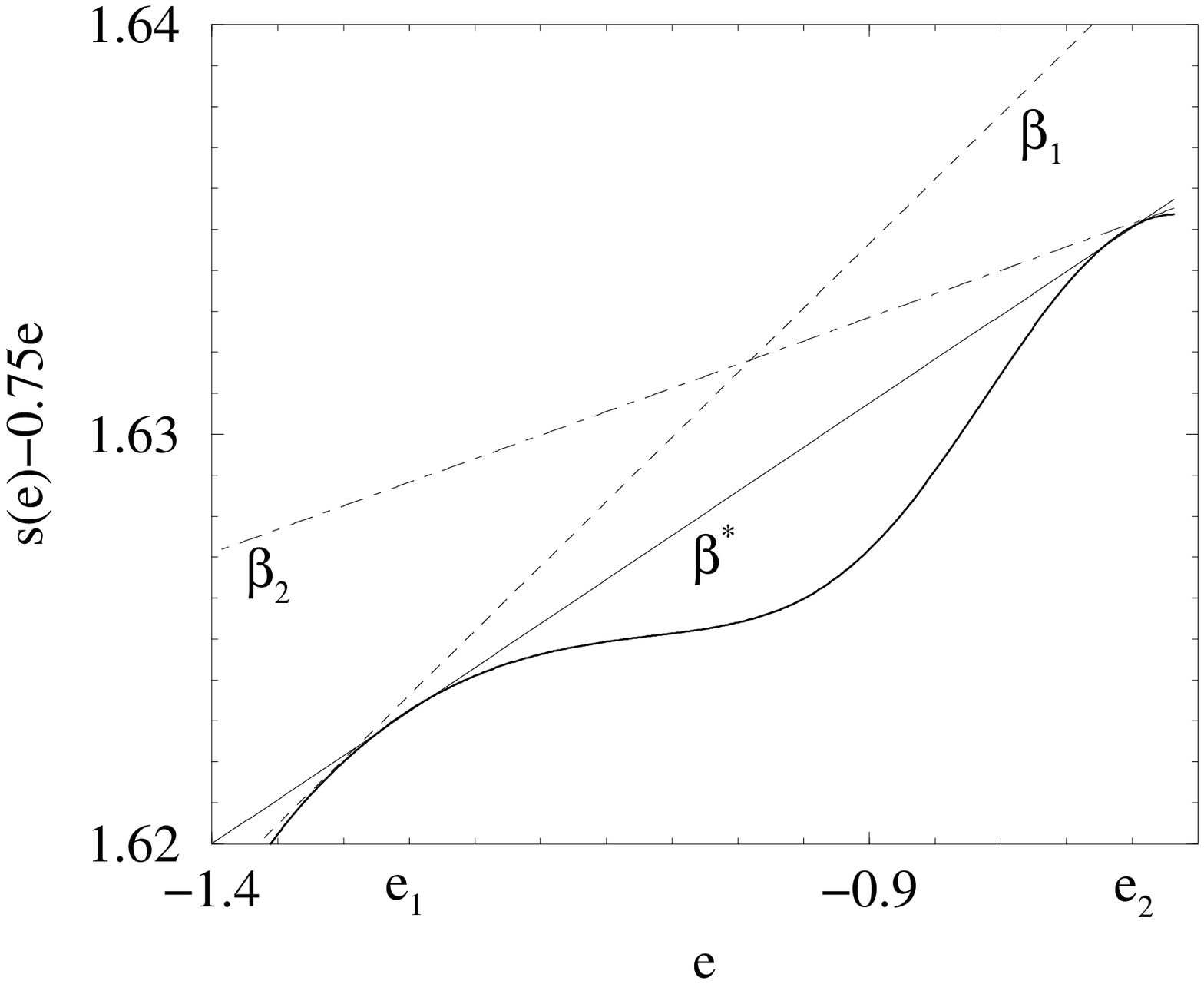}}
\noindent
{\small {\bf Fig.~8.}
Rescaled MFP entropy $s(e)-0.75e$ vs. energy 
$e$ for $q=3$ and $N=50$ (black solid line). 
Straight lines $\b e+\a$ have slopes:
$\b_1=0.7815,\; \b_2=0.7615$, and $\b^*=0.7715$.
}
\end{figure}

For $N\rightarrow \infty$, the integral (\ref{z}) can be evaluated
by the saddle point method which yields for the specific canonical
free energy $f(\b)$
\begin{equation}
\label{f}
f(\b)\equiv-{1\over \b N}\ln[Z(\b)]= e^* - {s(e^*)\over \b} +{\cal O}
({\ln N \over N}),
\end {equation}
where $e^*=e^*(\b)$ is the  solution of equation 
\begin{equation}
\label{e}
s'(e^*)=\b,
\end {equation}
that maximizes the expression in the exponential of (\ref{z}).
For a concave entropy $s(e)$, Eq.~(\ref{e}) has not more 
than one solution. For a system where the entropy has a 
convex dip, for a certain
$\b$, Eq.~(\ref{e}) may have two solutions, $e_1$ and $e_2$. 
This is the case
for all three $\b$ shown in Fig.~8. 
However, for both $\b_1$ and $\b_2$
a contribution from the second, smaller, maximum of
$s(e)-\b e$ to the integral (\ref{z}) is exponentially small in $N$.
For example, for $\b_1$ the left maximum $e_1$ is the global one, and
\begin{equation}
\label{su}
f(\b_1)\approx e_1 - {s(e_1)\over \b_1}- {e^{N[(s(e_2)-\b_1
e_2)-(s(e_1)-\b_1 e_1)]}
\over N}. 
\end {equation}
Only for $\b=\b^*$, do both maxima give equal contributions to $f(\b^*)$;
$\b^*$ can then be identified with the inverse temperature
of the first-order phase transition in the canonical ensemble.
It follows from (\ref{f},\ref{e}) that
\begin{equation}
\label{ph}
\left. {d[\b f(\b)]\over d\b}\right|_{\b\neq\b^*}=e^*,\;
 \left. {d[\b f(\b)]\over d\b}\right|_{\b=\b^*+0}=e_1,\;
 \left. {d[\b f(\b)]\over d\b}\right|_{\b=\b^*-0}=e_2, 
\end {equation}
which is in agreement with Ehrenfest's definition of a first-order
phase transition.

To put it another way, by performing the integration (\ref{f}) for various $\b$, 
we sample $s(e)$ for all $e$ except for those inside a convex dip,
and this sampling becomes a local one-to-one $e\rightarrow \b$ correspondence
for infinite $N$.
On the contrary, the contribution of a convex dip 
to $f(\b)$ is exponentially (in $N$) suppressed and vanishes completely for 
$N\rightarrow \infty$.

Now we have arrived to where we usually start from: a canonical ensemble
specific free energy $f(\b)$ calculated in the $N\rightarrow \infty$ 
limit. It is a continuous function of $\b$ with a jump discontinuity
in its first derivative at the first-order transition point $\b^*$.
As we showed above, it contains no information about the behavior of $s(e)$ 
in the interval
where $s(e)$ is convex. However, it is still worth finding out what kind
of microcanonical entropy can be  reconstructed from $f(\b)$.
In order to do that we
perform an inverse Laplace transformation of $f(\b)$:
\begin{equation}
\label{cm}
\tilde s (e)={1\over N} \ln \left \{{1\over 2\p i}\int_{\e-i\infty}^{\e+i\infty}
e^{N\b[e-f(\b)]} d\b\right \}
\end {equation}
In the limit of $N\rightarrow \infty$, we look for 
stationary points of the expression in the exponential,
$\b e -\b f(\b)$. Since for $\b\neq\b^*$, $d[\b f(\b)]/d\b=e^*(\b)$,
the extremum of the exponential 
is reached when $e^*(\b)=e$. Taking into account
(\ref{f},\ref{e}), we obtain that 
${\tilde s}(e)=s(e)$. This is true for
$e\leq e_1$ and $e\geq e_2$. Since $f(\b)$ has a cusp at $\b=\b^*$
and its left- and right- hand side derivatives are given by (\ref{ph}),
the derivatives of $\b[e-f(\b)]$ at $\b=\b^*$ for $e_1<e<e_2$ are:
\begin{equation}
\label{dr3}
 \left. {d[\b e - \b f(\b)]\over d\b}\right|_{\b=\b^*-0}=e-e_2<0,\;
 \left. {d[\b e - \b f(\b)]\over d\b}\right|_{\b=\b^*+0}=e-e_1>0. 
\end {equation}
Then for $e_1<e<e_2$, $\b^*$ gives the minimum of $\b[e-f(\b)]$
and in the saddle point approximation, integral (\ref{cm}) is reduced to
\begin{equation}
\label{sn}
{\tilde s} (e)= \b^*[e-f(\b^*)]+{\cal O}({\ln(N)\over N})
\end{equation}
From the definition of $\b^*$ (see Fig.~8) we have that
$\b^*(e_1-e_2)=s(e_1)-s(e_2)$ and $\b^* f(\b^*)=\b^* e_1 - s(e_1)$.
Substituting this into (\ref{sn}), one obtains for $e_1<e<e_2$:
\begin{equation}
\label{sb}
{\tilde s} (e)= {e-e_1 \over e_2-e_1}[s(e_2)-s(e_1)]+s(e_1)+
{\cal O}({\ln(N)\over N}) .
\end{equation}
So in a double transformed microcanonical entropy a convex dip
is replaced by a straight line (marked $\b^*$ in Fig.~8).

For any sufficiently smooth and not too rapidly growing
function, application of the direct 
and inverse Laplace transforms should leave this function unchanged.
However, exact canonical solutions are almost never available for 
finite $N$; more often they are known in the infinite system limit which
is singular.
As was shown above, if the $N\rightarrow \infty$ limit is taken 
before the integral (\ref{cm}) is evaluated, all information about
a possible convex intruder in $s(e)$ is totally lost.
Assume we study $f(\b)$ numerically for finite $N$.
Still, our chances to recover information about a convex dip in
$s(e)$ via the inverse Laplace transform of the CE partition function 
are quite low even for modest $N$
since
the contribution of  the convex dip to $f(\b)$ is 
exponentially small in $N$.
In statistical terms, it can be rephrased as  exponentially small 
accessibility
of low statistical weight states to canonical random sampling.

For extensive systems with first-order  phase transitions,
when the contribution of the interfacial states to $s(e)$ disappears
as $N\rightarrow \infty$, a convex intruders flattens into a
straight line. In this case a microcanonical behavior can be 
reconstructed from a known canonical solution via the procedure
outlined above. In fact, this reconstruction can be made operating 
solely with intensive quantities $f(\b)$ and $s(e)$ via
what can be called a generalized Legendre transform:
\begin{equation}
\label{lg}
{\tilde s} (e)= \b_e e - \b_e f(\b_e),\; \left. 
{d[\b f(\b)]\over d\b} \right|_{\b=\b_e}=e.
\end{equation}      
The second equation can only be solved for $e<e_1$ and $e>e_2$.
For $e_1<e<e_2$ these two branches of $s(e)$ are connected by 
a straight line (cf. (\ref{sb})).

\section{conclusion}
In two examples of the Potts model considered above, NN and MF, there is
a convex dip in the entropy plot when the system size is finite.
In the short-range-interacting NN model this dip flattens to a straight line
as the system size tends to infinity, while the dip in the entropy of the 
infinitely-long-interacting MF model persists in this limit.
This difference in behavior between NN and MF models is caused by
different relative entropic cost of interfacial states which scales
with the number of spins $N$ as $N^{d-1\over d}$ in the locally interacting 
models such as NN and goes to a constant in the MF model.
Since a convex dip in the microcanonical entropy 
is always associated with a negative specific heat interval, the
corresponding states in the canonical ensemble are thermodynamically 
unstable and are not represented in the canonical free energy.
Therefore in small or long-range interacting systems, 
these states of phase coexistence can only be studied
in the framework of the microcanonical ensemble.

\end{document}